\documentclass[%
 reprint,
 amsmath,amssymb,
 aps,
]{revtex4-1}

\usepackage{kpfonts}
\usepackage[T1]{fontenc}

\usepackage{enumerate}

\newcommand{\avg}[1]{\left< #1 \right>} 

\let\baraccent=\= 
\renewcommand{\=}[1]{\stackrel{#1}{=}} 

\usepackage{mathtools}

\newtheorem{thm}{Theorem}[section]

\usepackage{graphicx}
\usepackage{dcolumn}
\usepackage{bm}

\begin{document}

\preprint{APS/123-QED}

\title{The Total Green's Function of a Non-Interacting System}
\thanks{A footnote to the article title}%

\author{David Roberts}

\affiliation{%
 Jefferson Physical Laboratory, Harvard University, Cambridge, MA 02138\\}%

\date{\today}

\begin{abstract} \noindent Despite its centrality in the mathematical structure of perturbative many-body theory, the total Green's function for the many-body time-dependent Schrodinger equation has been ignored for decades, superseded by single-particle Green's functions, for which a vast portion of the literature has been devoted. In this paper, we give the first computation the total Green's function for the time-dependent Schrodinger equation for a non-interacting system of identical particles, setting the stage for a fresh interpretation of perturbative many-body physics.\end{abstract}

\pacs{Valid PACS appear here}
\maketitle



\section{\label{sec:level1}Introduction}

Condensed matter physicists use effective field theories all the time. In trying to engineer emergent  phenomena into novel materials at low-energy scales, following [2], it would be nice to be able to define a family of effective Hamiltonians 
\[\{H^\text{eff}[\Lambda]\}_{\Lambda\geq 0}\]
at every energy-scale $\Lambda$, for a fixed material, and formulate a renormalization group flow naturally in this context, that can interpolate between between a microscopic and effective many-body theory. As it turns out, the total Green's function 
\[G^\text{eff}[\Lambda]\equiv (\partial_t+iH^\text{eff}[\Lambda])^{-1}\]
is absolutely necessary to implement this program efficiently (See [1]). However, this puts us in an awkward position, because such an object is foreign to the condensed matter literature.

\section{\label{sec:level1}The Total Green's Function}
Time-evolution of a many-body quantum system is given by the time-dependent many-body Schrodinger equation (in units where $\hbar =1$), 
\[(\partial_t+iH)\Psi=0.\]
In this paper, we compute the Green's function of the linear differential equation above, which we'll call the {\bf total Green's function} of our many-body system:
\[G\equiv (\partial_t+iH)^{-1}.\]
Despite being such a fundamental mathematical quantity, surprisingly no one has actually computed the total Green's function for a many-body system.\\

In the case of a complicated interacting system, this computation is intractable. However, we can give the first computations of this function in the case that the dynamics is non-interacting. We can then compute the total Green's function of a general system via perturbation theory, but we will leave that for another article.\\

Since our Hamiltonian is non-interacting, it sends each $k$-particle portion of the total Hilbert space to itself: so the total Green's function $G$ also sends each $k$-particle portion of its Hilbert space to itself:
\begin{align*}
G&=\bigoplus_{k\geq 0}G_k
\end{align*}
The terms on the right-hand-side are the {\bf  k-particle Green's functions}. Therefore, to compute the total Green's function, it will suffice to compute $G_k$ for all $k\geq 0$.

\section{\label{sec:level1}Computation of $G_0$ and $G_1$}
{\bf Note}: For conceptual simplicity, throughout this article, we will assume that the single-particle Hilbert space is finite dimensional, with basis $\{f_i\}$. \\

The first two terms in the direct sum have already been computed in the literature, and we will not waste any time and just briefly mention the results here: 
\begin{align*}
(G_0)_{t}^{t'}&=(\partial_t +iH_0)^{-1})_{t}^{t'}
\end{align*}
Since $H_0\equiv 0$ (see [4]), we get the standard theta-function, the integral kernel of the differential operator $\partial_t$:
\begin{align*}
(G_0)_{t}^{t'}&=((\partial_t )^{-1})_t^{t'}=\theta(t-t')
\end{align*}
The operator $G_1$ is all over the many-body literature: it is called "the propagator", or sometimes "the single-particle Green's function". It has the following matrix elements:
\begin{align*}
(G_1)_{it}^{jt'}&=\theta(t-t')\avg{[a_\pm(f_i,t),a^\dag_\pm (f_j,t')]_\pm}_{T=0}
\end{align*}
Computation of $G_1$ is usually given as a trivial exercise in many-body textbooks, such as [3]. Already, with $k=0,1$, we can see a pattern forming. We will extrapolate to general values of $k$ in the next section.

\section{\label{sec:level1}Computation of $G_k$ for $k\geq 2$}
We now state our main result:
\begin{thm}[The Total Green's Function] For $k\geq 2$, define $\tilde{G}_k$ by the following matrix elements:
\begin{align*}
(\tilde{G}_k)_{i_1\cdots i_kt}^{j_1\cdots j_kt'}&=\theta(t-t')\big\langle[a_\pm(f_{i_1},t),a^\dag_\pm (f_{j_1},t')]_\pm\\
&\cdots [a_\pm(f_{i_k},t),a^\dag_\pm (f_{j_k},t')]_\pm\big\rangle_{T=0}
\end{align*}
Then the $k$-particle Green's function $G_k$ of the non-interacting system is simply the restriction of $\tilde{G}_k$ to the appropriate (anti-)symmetric subspace.
\end{thm}
{\it Proof.} For a system of non-interacting identical particles, there exists a basis of the single-particle Hilbert space with respect to which the Hamiltonian can be rewritten as
\begin{align*}
\sum_{ij}A_{ij}\,a_\pm^\dag(f_i) a_\pm(f_j)=\sum_{i}B_i\,a_\pm^\dag(g_i) a_\pm(g_i).
\end{align*}
{\bf Lemma}: {\it In the associated basis of our un-(anti)-symmetrized Fock-space, the matrix elements of $\tilde{G}_k$ become}
\begin{align*}
(\tilde{G}_k)_{i_1\cdots i_kt}^{j_1\cdots j_kt'}&=\theta(t-t')\big\langle[a_\pm(g_{i_1},t),a^\dag_\pm (g_{j_1},t')]_\pm\\
&\cdots [a_\pm(g_{i_k},t),a^\dag_\pm (g_{j_k},t')]_\pm\big\rangle_{T=0}
\end{align*}
{\it Proof of Lemma.} Let the change-of-basis be expressed as follows:
\begin{align*}
g_i=U_i^j f_j
\end{align*}
We now use the bilinearity of the (anti-)commutator, and the (anti-)linearity of $(a_\pm), a_\pm^\dag$:
\begin{align*}
(\tilde{G}_k)_{i_1\cdots i_kt}^{j_1\cdots j_kt'}&=\theta(t-t')\big\langle[a_\pm(g_{i_1},t),a^\dag_\pm (g_{j_1},t')]_\pm\\
&~~~~~~~~~~~~~~~~~~~~~~~~~\cdots [a_\pm(g_{i_k},t),a^\dag_\pm (g_{j_k},t')]_\pm\big\rangle_{T=0}\\
~~~~~~~~&=\theta(t-t')\big\langle U_{i_1}^{l_1}(U_{j_1}^{m_1})^*[a_\pm(f_{l_1},t),a^\dag_\pm (f_{m_1},t')]_\pm\\
&~~~~~~~~~~~~~~\cdots U_{i_k}^{l_k}(U_{j_k}^{m_k})^*[a_\pm(f_{l_k},t),a^\dag_\pm (f_{m_k},t')]_\pm\big\rangle_{T=0}\\
~~~~~~~~&=\theta(t-t')U_{i_1}^{l_1}(U_{j_1}^{m_1})^*\cdots U_{i_1}^{l_1}(U_{j_1}^{m_1})^*(\tilde{G}_k)_{l_1\cdots l_kt}^{m_1\cdots m_kt'}
\end{align*}
Since $U_i^j=(f_i,Uf_j)$ is unitary, we can write the above as follows:
\begin{align*}
(\tilde{G}_k)_{i_1\cdots i_kt}^{j_1\cdots j_kt'}&=(U^\dag)_{m_k}^{j_k}\cdots(U^\dag)_{m_1}^{j_1} (\tilde{G}_k)_{l_1\cdots l_kt}^{m_1\cdots m_kt'}U_{i_1}^{l_1}\cdots U_{i_k}^{l_k}
\end{align*}
Which satisfies the formula for induced change-of-basis on the un-(anti-)symmetrized Fock space $\square$.\\

Now we resume our proof. Since the Hamiltonian is diagonal in the basis $\{g_i\}$, it may be verified that 
\begin{align*}
a^\dag_\pm (g_j,t')=e^{-i(t-t')B_j}a_\pm^\dag(g_j,t)
\end{align*}
Therefore, we can begin to simplify the matrix elements of $\tilde{G}_k$ as follows:
\begin{align*}
(\tilde{G}_k)_{i_1\cdots i_kt}^{j_1\cdots j_kt'}&=\theta(t-t')\big\langle e^{-i(t-t')B_{j_1}}[a_\pm(g_{i_1},t),a^\dag_\pm (g_{j_1},t)]_\pm\\
&\cdots e^{-i(t-t')B_{j_k}}[a_\pm(g_{i_k},t),a^\dag_\pm (g_{j_k},t)]_\pm\big\rangle_{T=0}
\end{align*}
Using the equal-time (anti-)commutation relations 
\[\{a_\pm (f,t),a_\pm(g,t)\}=(f,g),\]
we get 
\begin{align*}
(\tilde{G}_k)_{i_1\cdots i_kt}^{j_1\cdots j_kt'}&=\theta(t-t')\delta_{i_1\cdots i_k}^{j_1\cdots j_k}\cdot e^{-i(t-t')(B_{j_1}+\cdots+B_{j_k})}.
\end{align*}
We now compute the time-derivative of the above expression: 
\begin{align*}
(\partial_t\tilde{G}_k)_{i_1\cdots i_kt}^{j_1\cdots j_kt'}&=\delta(t-t')\delta_{i_1\cdots i_k}^{j_1\cdots j_k}\cdot e^{-i(t-t')(B_{j_1}+\cdots+B_{j_k})}\\
&-i(B_{j_1}+\cdots+B_{j_k})(\tilde{G}_k)_{i_1\cdots i_kt}^{j_1\cdots j_kt'}
\end{align*}
Since the first term in the above expression vanishes for $t\neq t'$, we can eliminate the phase-factor which multiplies it, yielding
\begin{align*}
(\partial_t\tilde{G}_k)_{i_1\cdots i_kt}^{j_1\cdots j_kt'}&=\delta(t-t')\delta_{i_1\cdots i_k}^{j_1\cdots j_k}\\
&-i(B_{j_1}+\cdots+B_{j_k})(\tilde{G}_k)_{i_1\cdots i_kt}^{j_1\cdots j_kt'}.
\end{align*}
Recall that we can extend a non-interacting Hamiltonian to the un-(anti-)symmetrized Fock space by letting 
\[\tilde{H}_k(f_1\otimes \cdots \otimes f_k)\equiv \sum_{i}f_1\otimes\cdots \otimes H_1f_i\otimes \cdots \otimes f_k,\]
where $H_1$ is our associated single-particle Hamiltonian (see the first page for the definition of $H_1$). Therefore, acting on our Green's function with $i\tilde{H}_k$, we get
\begin{align*}
(i\tilde{H}_k\tilde{G}_k)_{i_1\cdots i_kt}^{j_1\cdots j_kt'}&=i(B_{j_1}+\cdots+B_{j_k})(\tilde{G}_k)_{i_1\cdots i_kt}^{j_1\cdots j_kt'}.
\end{align*}
Where we get the simple factor because we are implicitly in an eigenbasis of $\tilde{H}$, and so the action of $\tilde{H}$ is diagonal. Therefore, putting it all together, we get 
\begin{align*}
((\partial_t+i\tilde{H}_k)\circ (\tilde{G}_k))_{i_1\cdots i_kt}^{j_1\cdots j_kt'}=\delta(t-t')\delta_{i_1\cdots i_k}^{j_1\cdots j_k}
\end{align*}
In basis-independent language, this is the simple identity $(\partial_t+i\tilde{H}_k)\circ \tilde{G}_k=I$, i.e., we have verified that, on the un-(anti-)symmetrized Fock space, 
\begin{align*}
\tilde{G}_k=(\partial_t+i\tilde{H}_k)^{-1}.
\end{align*}
Therefore, restricting this operator expression to the (anti-)symmetrized Fock space $\mathcal F_\pm$ yields our desired identity:
\begin{align*}
~~~~~~~~~G_k=(\partial_t+iH_k)^{-1}.~~~~~\square
\end{align*}

\section{\label{sec:level1}Concluding Remarks}
In this paper, we computed the Green's function 
\[G=(\partial_t+iH)^{-1}\]
for the time-dependent Schrodinger equation, in the case of non-interacting identical particles, by computing each term in the direct-sum decomposition. The final result was $G= \bigoplus_{k\geq 0}P_\pm \tilde{G}_kP_\pm$, where $P_\pm$ is the (anti-)symmetrization operator, and
\begin{align*}
(\tilde{G}_k)_{i_1\cdots i_kt}^{j_1\cdots j_kt'}&=\theta(t-t')\big\langle[a_\pm(f_{i_1},t),a^\dag_\pm (f_{j_1},t')]_\pm\\
&\cdots [a_\pm(f_{i_k},t),a^\dag_\pm (f_{j_k},t')]_\pm\big\rangle_{T=0}.
\end{align*}
i.e., the $k$-particle Green's function $G_k$ of the non-interacting system is simply the restriction of $\tilde{G}_k$ to the appropriate (anti-)symmetric subspace.  {\bf Example}: {\it for a non-interacting system of identical spinless fermions, and in traditional notation,}
\begin{align*}
\tilde{G}_0(t,t')&=\theta(t-t')\\
\tilde{G_1}(x,x',t,t')&=\theta(t-t')\avg{\{\Psi(x,t),\Psi^\dag(x',t')\}}_{T=0}\\
G_2(x,x',y,y',t,t')&=\theta(t-t')\big\langle\{\Psi(x,t),\Psi^\dag(x',t')\}\\
&~~~~~~~~~~~~~~~~~~~~\cdot\{\Psi(y,t),\Psi^\dag(y',t')\}\big\rangle_{T=0}
\end{align*}
This work paves the way for a reformulation of perturbation and renormalization theory in terms of the full many-body Green's function.

\end{document}